%% file: paper.tex
\documentclass{article}
\usepackage{spconf,amsmath,graphicx}
\usepackage{bm}
\usepackage{pgfplots}
\usepackage{flushend}
\usepackage[ruled,vlined]{algorithm2e}
\usepackage{tikz}
\usepackage{multirow}
\usepackage{hyperref}
\usepackage{textcomp}
\usepackage{xcolor}
\DeclareUnicodeCharacter{2212}{−}
\usepgfplotslibrary{groupplots,dateplot}
\usetikzlibrary{patterns,shapes,arrows,arrows.meta,chains,fit,positioning,calc}
\pgfplotsset{compat=newest}
\usetikzlibrary{positioning}
\graphicspath{{figures/}}

\DeclareMathOperator*{\argmin}{\arg\!\min}

\newcommand{\fig}{Fig.\ }
\newcommand{\tab}{Table }

\def\distancefunction{{d}}
\def\distanceord{{o}}

\def\params{{\bm{\theta}}}

\def\featuresize{{M}}

\def\affinitycontrol{{\mu}}

\def\features{{\bm{f}}}
\def\featureselement{{f}}
\def\highresswir{{\bm{y}}}
\def\lowresswir{{\bm{s}}}

\def\affinitymatrix{{\bm{A}}}
\def\affinitymatrixelement{{}A}
\def\laplacianmatrix{{\bm{L}}}
\def\summatrix{{\bm{U}}}
\def\summatrixelement{{U}}
\def\downsamplematrix{{\bm{D}}}

\AtBeginShipout{\AtBeginShipoutUpperLeft{%
		\put(\dimexpr\paperwidth-20cm\relax,-0.5cm){\parbox[t]{19cm}{\copyright 2024 IEEE. Published in 2024 International Conference on Acoustics, Speech and Signal Processing (ICASSP), scheduled for 14-19 April 2024 in Seoul, Korea. Personal use of this material is permitted. However, permission to reprint/republish this material for advertising or promotional purposes or for creating new collective works for resale or redistribution to servers or lists, or to reuse any copyrighted component of this work in other works, must be obtained from the IEEE. DOI: \url{https://doi.org/10.1109/ICASSP48485.2024.10447935}}}%

}}
\title{A Guided Upsampling Network for Short Wave Infrared Images Using Graph Regularization}
\name{Frank Sippel, Jürgen Seiler, and André Kaup}
\address{Friedrich-Alexander-Universität Erlangen-Nürnberg\\
	Multimedia Communications and Signal Processing\\
	Cauerstraße 7, 91058 Erlangen, Germany\\
\thanks{The authors gratefully acknowledge that this work has been supported by
the Deutsche Forschungsgemeinschaft (DFG, German Research Foundation) under project number 491814627. \\
\noindent\hspace*{5mm}%
Source code and data: \textit{\url{https://github.com/FAU-LMS/gunet}}.
}}

\begin{document}

%
\maketitle
\begin{abstract}
	Exploiting the infrared area of the spectrum for classification problems is getting increasingly popular, because many materials have characteristic absorption bands in this area.
	However, sensors in the short wave infrared (SWIR) area and even higher wavelengths have a very low spatial resolution in comparison to classical cameras that operate in the visible wavelength area.
	Thus, in this paper an upsampling method for SWIR images guided by a visible image is presented.
	For that, the proposed guided upsampling network (GUNet) uses a graph-regularized optimization problem based on learned affinities is presented.
	The evaluation is based on a novel synthetic near-field visible-SWIR stereo database.
	Different guided upsampling methods are evaluated, which shows an improvement of nearly 1 dB on this database for the proposed upsampling method in comparison to the second best guided upsampling network.
	Furthermore, a visual example of an upsampled SWIR image of a real-world scene is depicted for showing real-world applicability.
\end{abstract}
\section{Introduction}
\label{sec:intro}

The infrared area of the spectrum is particularly interesting for a lot of real world classification problems, since different materials often have unique spectral fingerprints in this wavelength range.
For example, it can be used in agriculture~\cite{plant} to retrieve the plant health, in medicine to determine the degree of burn~\cite{burn}, in the area of recycling to discriminate between different types of plastic~\cite{plastics}, or in difficult imaging scenarios like haze, since longer wavelengths are advantageous in case of Rayleigh scattering~\cite{haze}.
Due to the different spectral area from roughly from 1000 nm to 2000 nm that is recorded by SWIR cameras, they typically have indium gallium arsenide (InGaAs) sensors.
Caused by the worse thermal conductivity of InGaAs sensors, the dark current noise is much stronger in comparison to classical visual range of spectrum cameras~\cite{swir_noise}.
Therefore, these cameras typically have huge pixel sizes and hence a much lower spatial resolution.
However, a high spatial resolution is highly desirable in many applications to reveal details and help classification and tracking algorithms.
Hence, the goal of the paper is to upsample the SWIR image by exploiting the structure of a corresponding high resolution visible image, which typically shows the spectral area from 400 nm to 700 nm.
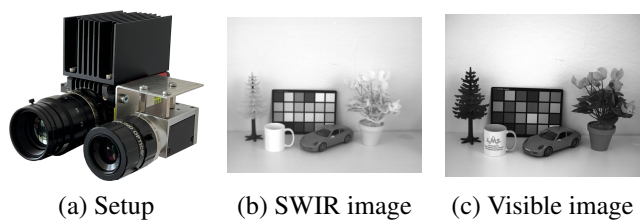
\begin{figure}
	\centering
	\input{figures/basic/basic.tikz}
	\vspace*{-0.5cm}
	\caption{The built stereo setup (a) producing an SWIR image (b) a visible image (c).}
	\label{fig:basic}
	\vspace*{-0.5cm}
\end{figure}

\begin{figure*}
	\centering
	\input{figures/gu/gu.tikz}
	\caption{The pipeline of the proposed guided upsampling network (GUNet).}
	\label{fig:gu}
	\vspace*{-0.3cm}
\end{figure*}
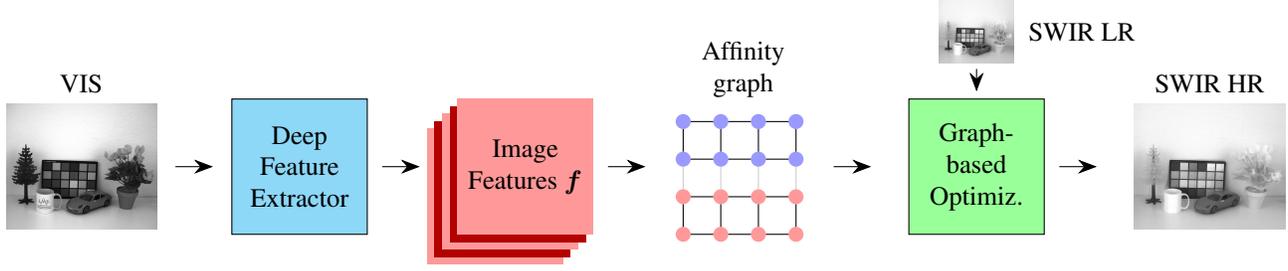

To test the real-world guided upsampling capability, a visible-SWIR stereo camera setup was built as shown in \fig\ref{fig:basic}.
The SWIR camera has a resolution of 320 \texttimes\ 256 pixels with a pixel size of 30 \textmu m \texttimes\ 30 \textmu m, while the visible camera has a resolution of 2448 \texttimes\ 2048 using pixels of size 3.45 \textmu m \texttimes\ 3.45 \textmu m.
The lenses of both cameras have a focal length of 16 mm.
Due to the slightly bigger sensor size, the SWIR camera is able to capture a wider field of view.

The visible-SWIR stereo setup shown in \fig\ref{fig:basic} operates in near-field, which means that the objects are relatively close in comparison to, e.g., an airborne device.
In near-field imaging, the objects in the scene have a depth-dependent offset, also called disparity, from the perspective of spatially distributed cameras.
Moreover, due to the different viewing angles on objects, the different cameras also have a different perspective on objects and see different parts of the background behind an object.
Hence, a registration and reconstruction process is necessary.
For this, an image processing pipeline like introduced by Genser et al.~\cite{camsi} can be used.
The assumption for the proposed method is that such a pipeline was already deployed.

The task of guided upsampling is popular for increasing the resolution of sparse depth maps based on a high resolution image.
There, the goal is to upsample a sparse depth map using a high resolution RGB image as guide~\cite{gad, fdkn, graph_sr}.
In this paper, the idea of using affinities~\cite{graph_sr} is examined, since SWIR images typically have less texture than its corresponding visible image.
This is visible in \fig\ref{fig:basic}, where the logo of the cup is not visible in the SWIR image.
Moreover, an affinity-based upsampling is more robust towards registration and reconstruction errors made by the image processing pipeline for overlaying the SWIR and visible image.

\section{Guided Upsampling}
\label{sec:guided_upsampling}

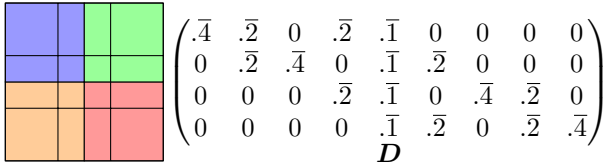
\begin{figure}
	\centering
	\input{figures/downsampling/downsampling.tikz}
	\caption{Illustration of the downsampling matrix $\downsamplematrix$. A high resolution $3 \times 3$ grid is downscaled to a $2 \times 2$ image. The low resolution pixels are shown in different colors.}
	\label{fig:downsampling}
	\vspace*{-0.2cm}
\end{figure}

The proposed guided upsampling network (GUNet) is based on a work for guided depth upsampling by Lutio et al.~\cite{graph_sr}.
In this paper, this network is improved and modified to upsample SWIR image data using the visible image as guide.
For that, a more general affinity function is introduced and the optimal working point is examined.
Moreover, arbitrary scaling factors are made possible by introducing a more general downsampling operator.
Finally, it is proven that the optimization problem can be backpropagated, and thus the network is end-to-end trainable.
It is assumed that the images are already registered and reconstructed using e.g.~\cite{camsi}.

\subsection{Network Architecture}

As depicted in \fig\ref{fig:gu}, the first step is to extract pixel-wise deep features $\features$ of length $\featuresize$ from the visible image using a UNet~\cite{unet} architecture, namely ResNet50~\cite{resnet}.
In this paper, the length $\featuresize$ of the features is set to 64.
With the help of these features, the affinity matrix $\affinitymatrix$, which contains the affinity for each pixel to its four direct neighbors, can be calculated by the similarity function
\begin{equation}
	\affinitymatrixelement_{ij} = \mathrm{e}^{-\frac{\distancefunction(\featureselement_i, \featureselement_j)}{\affinitycontrol}},
\end{equation}
where $\affinitycontrol$ is a learnable scaling parameter and $\distancefunction(\featureselement_i, \featureselement_j)$ is a distance function between features.
This distance function is discussed in Section~\ref{sec:affinity}.
$\affinitymatrix$ can be interpreted as adjacency matrix of an affinity graph.
Afterwards, the Laplacian matrix $\laplacianmatrix = \summatrix - \affinitymatrix$ is calculated, where the degree matrix $\summatrix$ is a diagonal matrix with the entries being $\summatrixelement_{ii} = \sum_{j} \affinitymatrixelement_{ij}$.
Finally, the optimization problem
\begin{equation}
	\label{eq:opt}
	\argmin_{\highresswir} \mathcal{L}_{\text{rec}}(\highresswir, \laplacianmatrix) = \argmin_{\highresswir} ||\downsamplematrix\highresswir - \lowresswir||_2^2 + \lambda \highresswir^{\text{T}}\laplacianmatrix\highresswir
\end{equation}
is solved, where the first term is the data term and contains the downsampling matrix $\downsamplematrix$, the vectorized low resolution SWIR image $\lowresswir$ and the vectorized high resolution SWIR image $\highresswir$ to determine.
The second term is the regularizer, which implicitly incorporates the affinity matrix and thus the structure of the high resolution visible image.
The form of this term results from any smoothness regularizer.
In this case, the estimated signal should be smooth on the affinity graph.
$\lambda$ is a learnable trade-off parameter steering how close to stay to the original image.
Since the scale factor between warped SWIR image and the visible image is non-integer with a very high probability, $\downsamplematrix$ is able to split a high resolution pixel to influence several low resolution pixels.
For example, for a scale factor of $1.5$, the first low resolution pixel will contain the content of the first high resolution pixel, half of the pixels to the right and to the bottom, and a quarter of the diagonal pixel.
This example is depicted in \fig\ref{fig:downsampling}.
The solution to the optimization problem is calculated by solving the linear systems of equations
\begin{equation}
	\label{eq:solution}
	\left( \downsamplematrix^{\text{T}}\downsamplematrix + \lambda \laplacianmatrix \right) \highresswir = \downsamplematrix^{\text{T}}\lowresswir.
\end{equation}

The optimization problem shown in \eqref{eq:opt} can be backpropagated.
The gradients through the optimization problem can be calculated using the implicit function theorem~\cite{implicit_function_theorem}.
In general, the optimization problem for training contains the optimization problem shown in~\eqref{eq:opt}
\begin{equation}
	\params^{*} = \argmin_{\params} \mathcal{L}_{\text{train}} \left(  \argmin_{\highresswir} \mathcal{L}_{\text{rec}}(\highresswir, \laplacianmatrix(\params)) \right),
\end{equation}
where it is being made explicit, that only the Laplacian matrix depends on the network parameters $\laplacianmatrix$.
The gradients for this type of loss can be calculated using the implicit function theorem.
Suppose $\laplacianmatrix$ and $\highresswir$ are related to each other through function $g$
\begin{equation}
	\label{eq:ift}
	g(\highresswir, \laplacianmatrix) = 0.
\end{equation}
Then, by assuming $g(\highresswir, \laplacianmatrix)$ to be smooth, if $\highresswir$ is slightly changed by $\Delta_\highresswir$, $\laplacianmatrix$ is also slightly changed by $\Delta_\laplacianmatrix$ to preserve the constraint
\begin{equation}
	g(\Delta_\highresswir, \Delta_\laplacianmatrix) = 0.
\end{equation}
Taking the first-order expansion of this leads to
\begin{equation}
	g(\highresswir, \laplacianmatrix) + \Delta_\highresswir \frac{\delta g}{\delta \highresswir} + \Delta_\laplacianmatrix \frac{\delta g}{\delta \laplacianmatrix} = 0.
\end{equation}
Since \eqref{eq:ift} holds, the relationship
\begin{equation}
	\Delta_\highresswir \frac{\delta g}{\delta \highresswir} = -\Delta_\laplacianmatrix \frac{\delta g}{\delta \laplacianmatrix}
\end{equation}
is established.
Finally,
\begin{equation}
	\frac{\Delta_\highresswir}{\Delta_\laplacianmatrix} = -\left(\frac{\delta g}{\delta \highresswir}\right)^{-1} \frac{\delta g}{\delta \laplacianmatrix}
\end{equation}
holds.
Here, the function $g(\highresswir, \laplacianmatrix) = \frac{\delta \mathcal{L}_{\text{rec}}(\highresswir, \laplacianmatrix)}{\delta\highresswir}$ is fulfilled, which leads to
\begin{equation}
	\begin{aligned}
	\frac{\delta \highresswir}{\delta \laplacianmatrix} &= -\left(\frac{\delta^2 \mathcal{L}_{\text{rec}}(\highresswir, \laplacianmatrix)}{\delta \highresswir \delta \highresswir^{\text{T}}}\right)^{-1} \frac{\delta^2 \mathcal{L}_{\text{rec}}(\highresswir, \laplacianmatrix)}{\delta \highresswir \delta \laplacianmatrix^{\text{T}}} \\
		&= -\left( \downsamplematrix^{\text{T}}\downsamplematrix + \lambda \laplacianmatrix \right)^{-1} \lambda \highresswir
	\end{aligned}
\end{equation}
Instead of calculating the inverse directly, a linear system of equations $\left( \downsamplematrix^{\text{T}}\downsamplematrix + \lambda \laplacianmatrix \right)\bm{x} = \lambda \highresswir$ is solved very similar to \eqref{eq:solution}.
Now, this gradient can be used in the chain rule to get the gradient of the training loss function with respect to the network parameters
\begin{equation}
	\frac{\delta \mathcal{L}_{\text{train}}}{\delta \laplacianmatrix} = \frac{\delta \mathcal{L}_{\text{train}}}{\delta \highresswir} \frac{\delta \highresswir}{\delta \laplacianmatrix}.
\end{equation}
This shows that the whole network is end-to-end trainable.

The problem is that no large scale visible-SWIR training data is available.
Therefore, the RGB database Places~\cite{places} is used and the cross spectral data is augmented.
For this, the same data augmentation is used as in~\cite{dgnet}.
This data augmentation exploits the HSV color space to assign random grayscale values to a couple of colors.
In between these colors, the remaining colors are linearly interpolated using the random grayscale values.
Patches of size $256 \times 256$ were used as well as a downscale factor of 8.
Adam optimizer with parameters $\beta_1 = 0.5$ and $\beta_2 = 0.999$ and a learning rate of $0.0001$ was used.
The best model on training data after one epoch is deployed.

\subsection{Affinity Function}
\label{sec:affinity}
\begin{table}[t]
	\centering
	\caption{Average PSNR and SSIM of different distance functions $d_\distanceord(\featureselement_i, \featureselement_j)$ on the hyperspectral database CAVE.}
	\vspace*{0.2cm}
	\label{tab:ord}
	\begin{tabular}{c|ccccc}
						& $o=1$ & $o=1.5$ & $o=2$ & $o=4$ & $o=10$ \\
		\hline
				PSNR & 32.13 & \textbf{32.32} & 32.02 & 32.25 & 31.88\\

				SSIM & 0.890 & \textbf{0.892} & 0.889 & 0.889 & 0.880\\
	\end{tabular}
\end{table}

One of the very few choices to make in this architecture is the affinity and thus distance function.
For this, the network was trained using several distance functions.
The distance functions have the general form of
\begin{equation}
	d_\distanceord(\featureselement_i, \featureselement_j) = \frac{1}{\featuresize} \sum_{m=1}^{\featuresize} |\featureselement_{i,m} - \featureselement_{j,m}|^{\distanceord}
\end{equation}
and were used at points $\distanceord\in\{1, 1.5, 2, 4, 10\}$.

To optimize this parameter, the individual trained networks were evaluated on the hyperspectral database CAVE~\cite{cave}.
For this an image at wavelength 500 nm served as guide while the image at wavelength 650 nm was scaled down by factor 8 and used as low resolution image.
Note that the reconstruction and registration process necessary for visible-SWIR stereo imaging is not part of the evaluation here.
Only the upsampling performance is considered for finding the best distance function.

The results are summarized in \tab\ref{tab:ord}.
The different distance functions $d_\distanceord(\featureselement_i, \featureselement_j)$ all work well on the CAVE database.
However, parameters $\distanceord = 1.5$ and $\distanceord = 4$ work better than the others.
Due to the slightly better performance, $\distanceord = 1.5$ is chosen in the upcoming evaluation.

\section{Evaluation}
\label{sec:evaluation}

The evaluation is split into two parts.
First, a quantitative evaluation based on a novel visible-SWIR database is shown.
Afterwards, a qualitative evaluation is performed using a record from a real-world visible-SWIR stereo setup.

\subsection{Quantitative Evaluation}

\begin{figure}
	\centering
	\input{figures/hyvid/hyvid.tikz}
	\vspace*{-0.2cm}
	\caption{The novel synthetic visible-SWIR stereo database. The image on the left was rendered from the low-resolution SWIR camera, while the right image depicts the high resolution visible image.}
	\label{fig:hyvid}
	\vspace*{-0.2cm}
\end{figure}
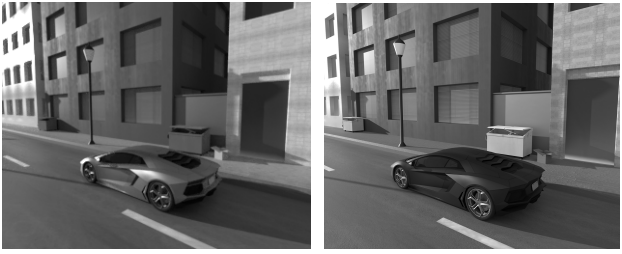

Since it is impossible to record ground-truth data using the presented setup, a synthetic visible-SWIR database is created based on the database HyViD \cite{hyvid}, which contains wavelengths from 400 nm to 700 nm.
This database was created using the 3D modeling software Blender.
Textures were extracted from a real-world database containing scenes from different environments.
Light sources were emulated using Planck's law.
The database contains seven moving scenes, each with 30 frames.
For this evaluation, a visible camera and a SWIR camera in a stereo setup was synthetically created and inserted into the scenes.
The visible camera was rendered at 500 nm, while the SWIR camera was rendered at 650 nm, since no SWIR textures are available.
Apart from the focal length (6 mm), all other aspects stayed as close to the real setup as possible.
An example frame from both cameras of the scene \textit{city} is shown in \fig\ref{fig:hyvid}.

Note that the cameras are perfectly aligned in Blender, and thus the quantitative evaluation skips the calibration process.
Instead, since the SWIR camera has a slightly bigger field of view, the parts of the sensor, which are not visible to the visible camera, are cropped away.
These areas were identified by comparing physical sensor sizes.

\begin{table}[t]
	\centering
	\vspace*{-0.2cm}
	\caption{Average PSNR in dB and SSIM of different upsampling methods on the novel synthetic database.}
	\vspace*{0.2cm}
	\label{tab:metrics}
	\begin{tabular}{@{\hspace*{0.0cm}}c@{\hspace*{0.0cm}}|@{\hspace*{0.1cm}}c@{\hspace*{0.1cm}}c@{\hspace*{0.1cm}}c@{\hspace*{0.1cm}}c@{\hspace*{0.1cm}}c@{\hspace*{0.1cm}}c@{\hspace*{0.1cm}}c@{\hspace*{0.1cm}}c@{}}
						& BIC & VDSR & HAN & SwinIR & GAD & FDKN & GUNet \\
						& & \cite{vdsr} & \cite{han} & \cite{swinir} & \cite{gad} & \cite{fdkn} & \\
		\hline
				PSNR & 25.15 & 25.14 & 24.51 & 25.03 & 23.83 & 26.40 & \textbf{27.31}\\

				SSIM & 0.755 & 0.754 & 0.629 & 0.754 & 0.745 & 0.820 & \textbf{0.834}\\
	\end{tabular}
	\vspace*{-0.2cm}
\end{table}

GUNet is compared against single image super resolution methods bicubic interpolation (BIC), Very Deep Super Resolution~\cite{vdsr} (VDSR), Holistic Attention Network~\cite{han} (HAN) and Swin Image Restoration~\cite{swinir} (SwinIR), as well as guided upsampling methods Guided Anisotropic Diffusion~\cite{gad} (GAD), and Fast Deformable Kernel Networks~\cite{fdkn} (FDKN).
The guided upsampling networks GAD and FDKN were retrained using the same procedure as GUNet, since they all originate from guided depth upsampling.

The results in terms of average PSNR and SSIM are summarized in \tab\ref{tab:metrics}.
The single image super resolution methods are able to upsample the SWIR image, but cannot keep up with the performance of the guided methods according to PSNR and SSIM.
Surprisingly, the bicubic upsampled version performs best in the group of single-image super resolution methods.
This may originate from the fact, that these methods were trained on only downsampled versions of the images to upsample.
Hence, registration and reconstruction errors are interpreted as details that needs to be sharpened and thus overexaggerated, while BIC rather smoothens these errors.
A similar effect is happening with the guided diffusion-based network GAD, which is performing much worse than just a simple bicubic upscaling.
On the other hand, FDKN and GUNet are able to properly outperform the bicubic upscaled version.
The proposed GUNet performs best according to PSNR and SSIM.

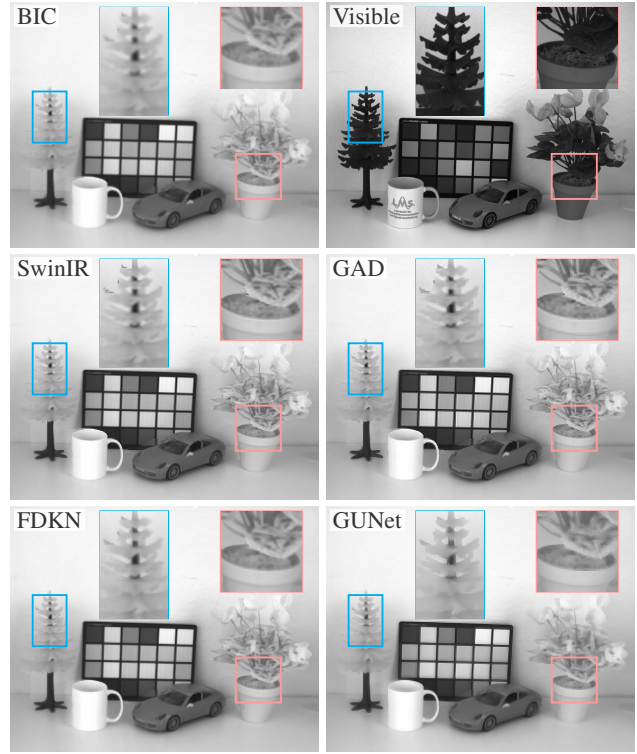
\begin{figure}
	\centering
	\input{figures/evaluation/evaluation.tikz}
	\caption{The qualitative evaluation on a real-world scene.}
	\label{fig:evaluation}
	\vspace*{-0.3cm}
\end{figure}

\vspace*{-0.2cm}

\subsection{Qualitative Evaluation}

For the qualitative evaluation, a real-world scene was recorded using the presented setup in \fig\ref{fig:basic}.
In \fig\ref{fig:evaluation}, the results of the bicubic upsampled image, the visible guide image, various upsampling methods are depicted.
Due to the affinity-based optimization problem, GUNet provides sharper edges than all other methods.
This is well visible for the tree trunk, the color calibration chart and the plant.
Note that GUNet can even better conceal some of the reconstruction errors made by the cross spectral reconstruction, e.g., to the left of the car.

\vspace*{-0.2cm}

\section{Conclusion}
\label{sec:conclusion}

\vspace*{-0.2cm}

In this paper, a novel near-field visible-SWIR stereo camera setup was introduced.
Due to unfavorable thermal properties, the SWIR sensor has a much lower resolution than the visible sensor.
Therefore, a guided upsampling network using a graph-regularized optimization problem was presented exploiting the high resolution visible image.
In the evaluation, the proposed affinity-based network outperformed its learned competitors trained with the same procedure by 1 dB and gains 2 dB over bicubic upscaling.
Moreover, this guided upsampling method also provides satisfying results on a real-world visible-SWIR record.

\vfill\pagebreak\clearpage

\bibliographystyle{IEEEbib}
\bibliography{refs}

\end{document}

%% file: figures/basic/basic.tikz
\begin{tikzpicture}[y=-1cm]
    \node (cam) at (0, 0) {\includegraphics[width=0.3\linewidth]{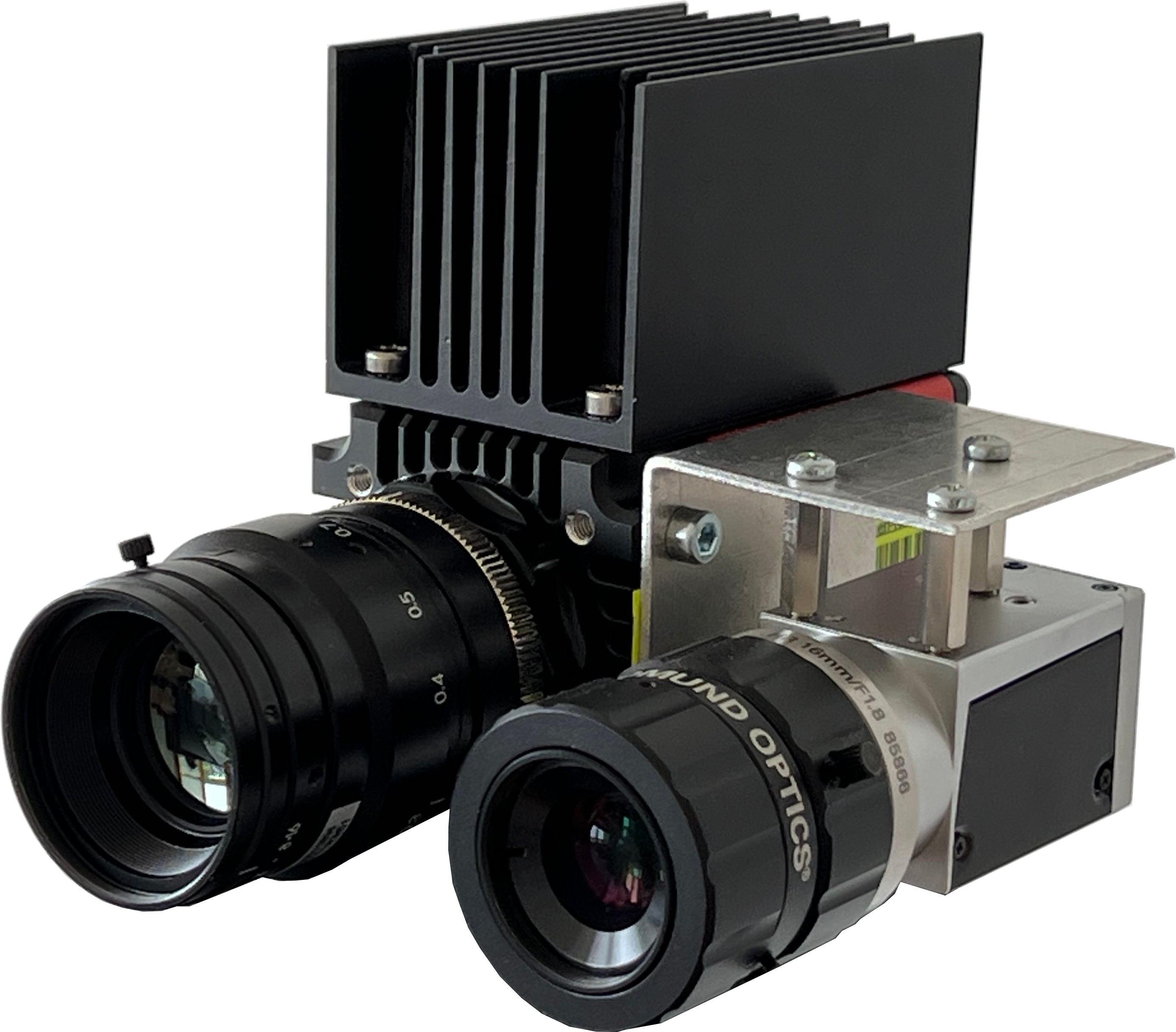}};
    \node[inner sep=0pt, align=center] (a) at (0, 1.5) {(a) Setup\strut};
    \node (cam) at (2.9, 0) {\includegraphics[width=0.3\linewidth]{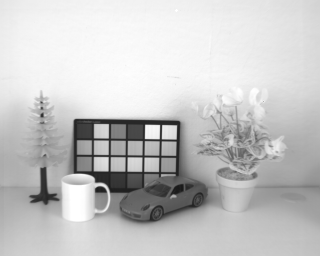}};
    \node[inner sep=0pt, align=center] (a) at (2.9, 1.5) {(b) SWIR image\strut};
    \node (cam) at (5.8, 0) {\includegraphics[width=0.3\linewidth]{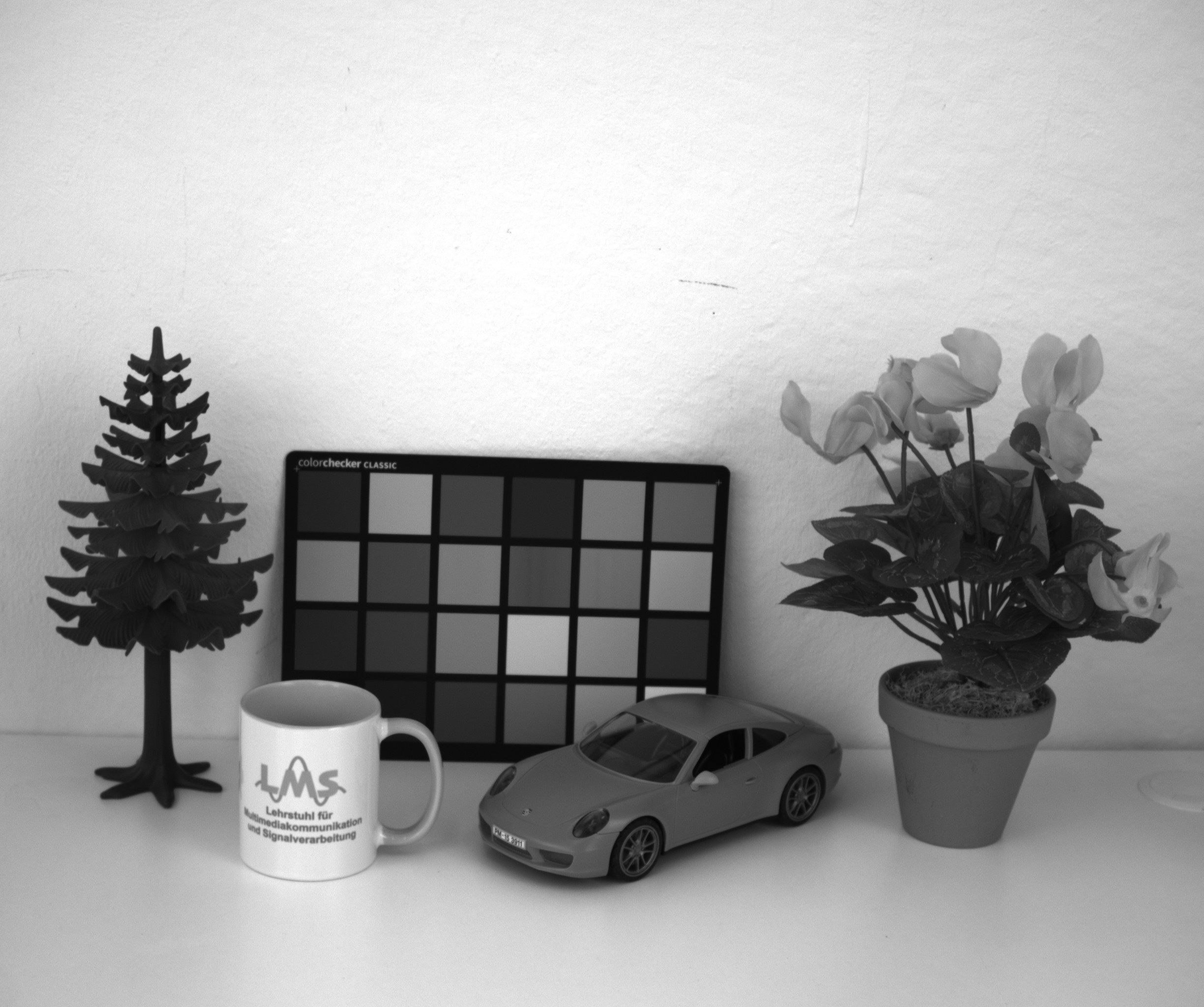}};
    \node[inner sep=0pt, align=center] (a) at (5.8, 1.5) {(c) Visible image\strut};
\end{tikzpicture}

%% file: figures/gu/gu.tikz
\begin{tikzpicture}
    \node[inner sep=0pt] (guide) at (0,0) {\includegraphics[width=2cm]{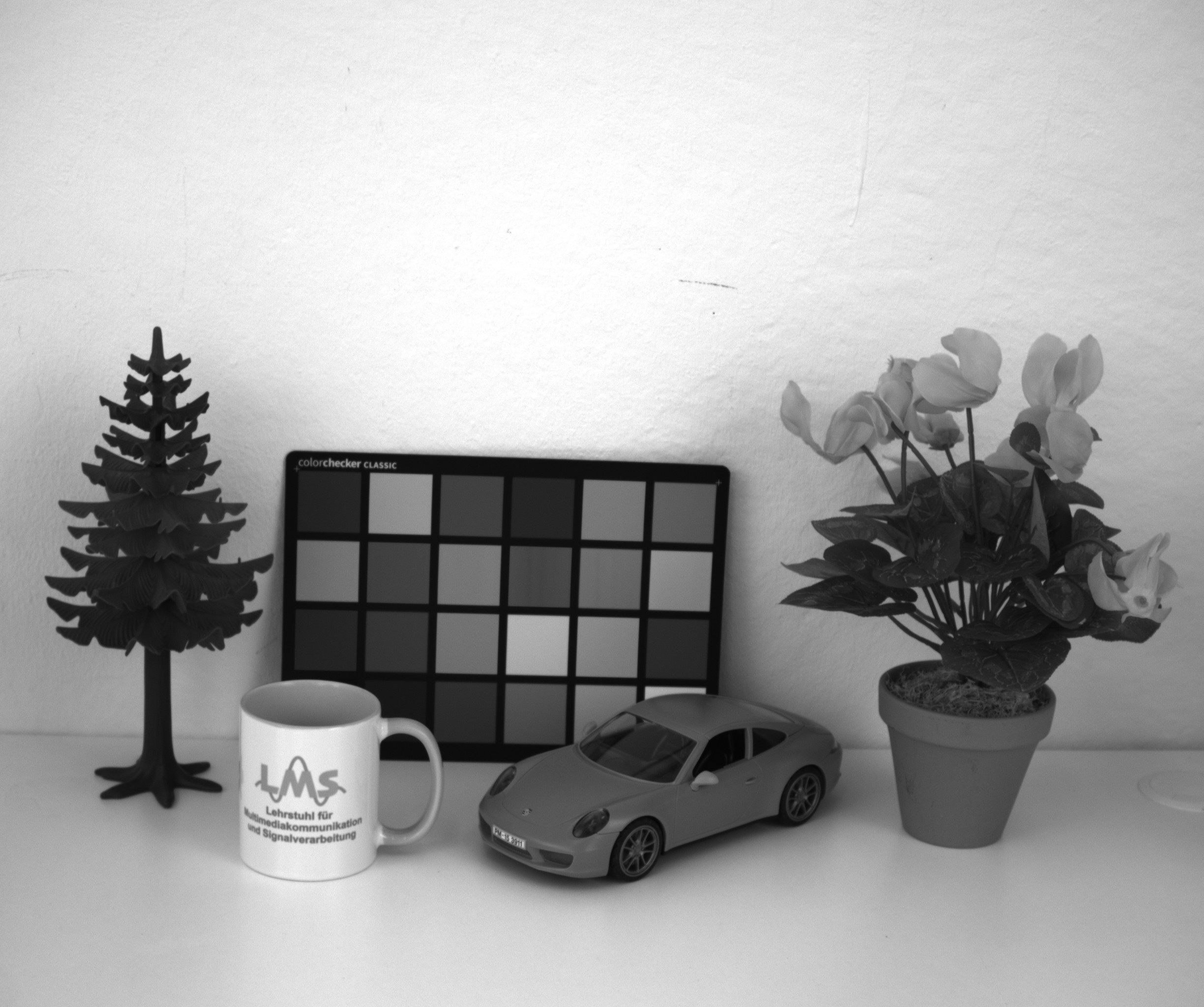}};
    \node[inner sep=0pt, align=center] (lef) at (0, 1.1) {VIS};

    \draw[fill=white!60!cyan, draw=black] (2,-0.9) rectangle ++(1.8, 1.8) node[pos=0.5, text=black, align=center] (features) {Deep\\Feature\\Extractor};

    \draw[-{Stealth[scale=1.5]}] (4, 0) -- (4.5, 0);
    \draw[-{Stealth[scale=1.5]}] (1.25, 0) -- (1.75, 0);

    \begin{scope}[shift={(5, -0.9)}]
        \fill[fill=white!60!red] (-0.4, -0.4) rectangle ++(1.8, 1.8);
        \fill[fill=black!30!red] (-0.3, -0.3) rectangle ++(1.8, 1.8);
        \fill[fill=white!60!red] (-0.2, -0.2) rectangle ++(1.8, 1.8);
        \fill[fill=black!30!red] (-0.1, -0.1) rectangle ++(1.8, 1.8);
        \fill[fill=white!60!red] (0, 0) rectangle ++(1.8, 1.8) node[pos=0.5, text=black, align=center] (features) {Image\\Features $\features$};
    \end{scope}

    \draw[-{Stealth[scale=1.5]}] (7, 0) -- (7.5, 0);

    \begin{scope}[shift={(8, -0.9)}]
        \foreach \x in{0,...,3}
        {
            \fill[fill=white!60!red] (0 + \x * 0.5, 0) circle (0.1);
            \fill[fill=white!60!red] (0 + \x * 0.5, 0.5) circle (0.1);
        }
        \foreach \x in{0,...,3}
        {
            \fill[fill=white!60!blue] (0 + \x * 0.5, 1.0) circle (0.1);
            \fill[fill=white!60!blue] (0 + \x * 0.5, 1.5) circle (0.1);
        }
        \foreach \x in{0,...,2}
        {
            \foreach \y in{0,...,3}
            {
                \draw[opacity=1] (\x * 0.5 + 0.1, \y * 0.5) -- (\x * 0.5 + 0.4, \y * 0.5);
            }
        }
        \foreach \x in{0,...,3}
        {
            \draw[opacity=1] (\x * 0.5, 0 * 0.5 + 0.1) -- (\x * 0.5, 0 * 0.5 + 0.4);
            \draw[opacity=0.2] (\x * 0.5, 1 * 0.5 + 0.1) -- (\x * 0.5, 1 * 0.5 + 0.4);
            \draw[opacity=1] (\x * 0.5, 2 * 0.5 + 0.1) -- (\x * 0.5, 2 * 0.5 + 0.4);
        }
        \node[inner sep=0pt, align=center] (lef) at (0.8,2.2) {Affinity\\graph};
    \end{scope}

    \draw[-{Stealth[scale=1.5]}] (10, 0) -- (10.5, 0);

    \draw[fill=white!60!green, draw=black] (11,-0.9) rectangle ++(1.8, 1.8) node[pos=0.5, text=black, align=center] (features) {Graph-\\based\\Optimiz.};
    \node[inner sep=0pt, align=center] (lef) at (13.3, 1.8) {SWIR LR};
    \node[inner sep=0pt] (lr) at (11.9,1.8) {\includegraphics[width=1cm]{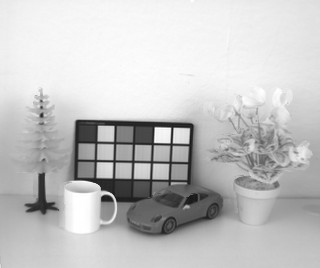}};

    \draw[-{Stealth[scale=1.5]}] (13, 0) -- (13.5, 0);
    \draw[-{Stealth[scale=1.5]}] (11.9, 1.3) -- (11.9, 1);

    \node[inner sep=0pt, align=center] (lef) at (15, 1.1) {SWIR HR};
    \node[inner sep=0pt] (lr) at (15,0) {\includegraphics[width=2cm]{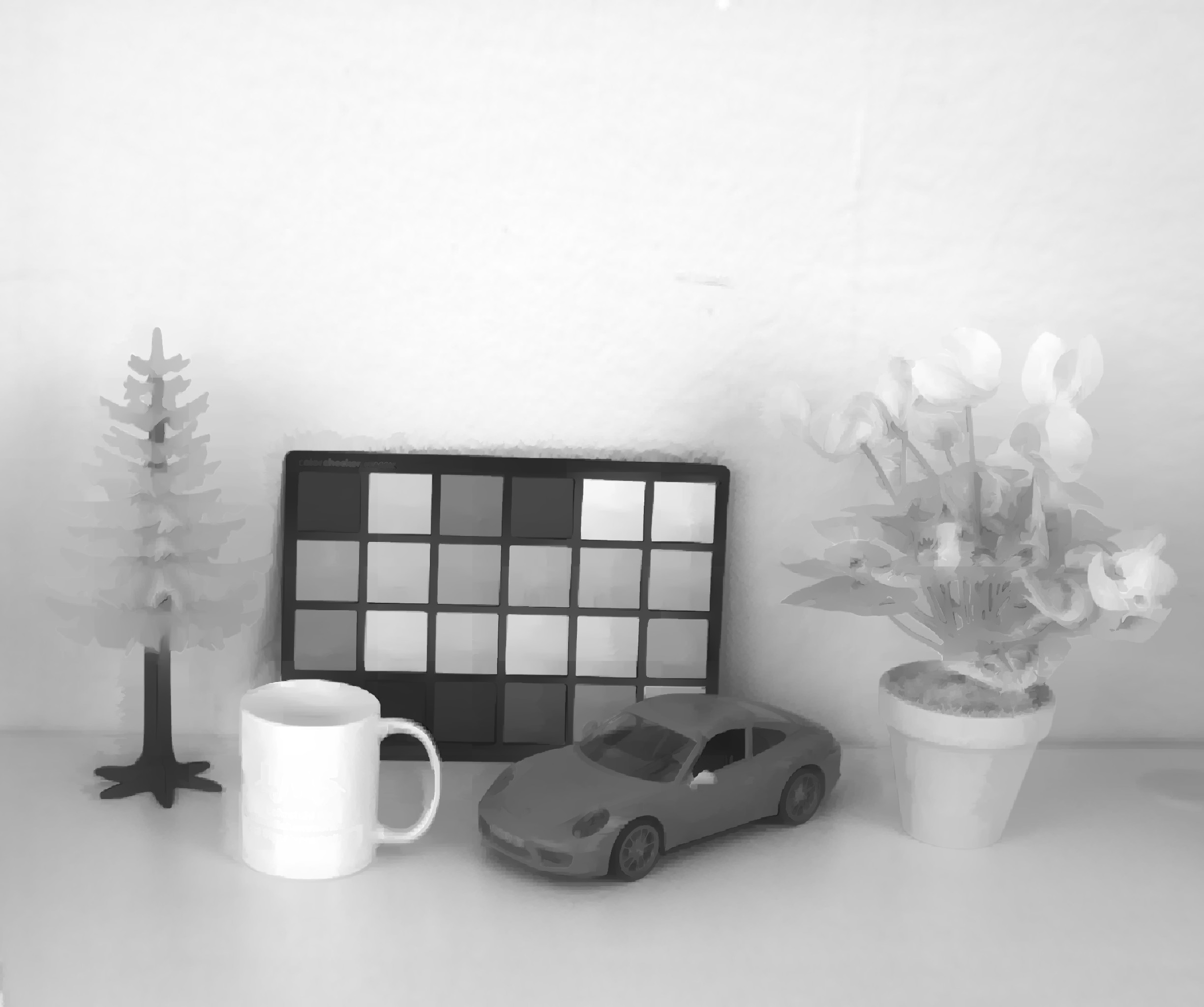}};
\end{tikzpicture}

%% file: figures/downsampling/downsampling.tikz
\begin{tikzpicture}
    \draw[fill=white!60!orange] (0,0) rectangle ++(1.05, 1.05);
    \draw[fill=white!60!red] (1.05,0) rectangle ++(1.05, 1.05);
    \draw[fill=white!60!blue] (0,1.05) rectangle ++(1.05, 1.05);
    \draw[fill=white!60!green] (1.05,1.05) rectangle ++(1.05, 1.05);
    \draw[step=0.7,black,thin] (0,0) grid (2.1, 2.1);

    \node[inner sep=0pt, align=center] (lef) at (5.1, 1.05) {$
    \begin{pmatrix}
        .\overline{4} & .\overline{2} & 0 & .\overline{2} & .\overline{1} & 0 & 0 & 0 & 0\\
        0 & .\overline{2} & .\overline{4} & 0 & .\overline{1} & .\overline{2} & 0 & 0 & 0\\
        0 & 0 & 0 & .\overline{2} & .\overline{1} & 0 & .\overline{4} & .\overline{2} & 0\\
        0 & 0 & 0 & 0 & .\overline{1} & .\overline{2} & 0 & .\overline{2} & .\overline{4}
    \end{pmatrix}
    $};
    \node[inner sep=0pt, align=center] (lef) at (5.1, 0.1) {$\downsamplematrix$};
\end{tikzpicture}

%% file: figures/hyvid/hyvid.tikz
\begin{tikzpicture}
    \node (swir) at (0, 0) {\includegraphics[width=.23\textwidth]{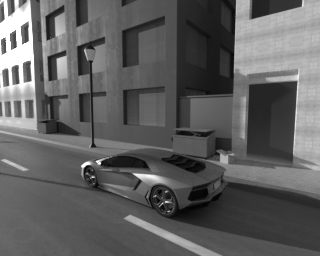}};
    \node (vis) at (4.2, 0) {\includegraphics[width=.22\textwidth]{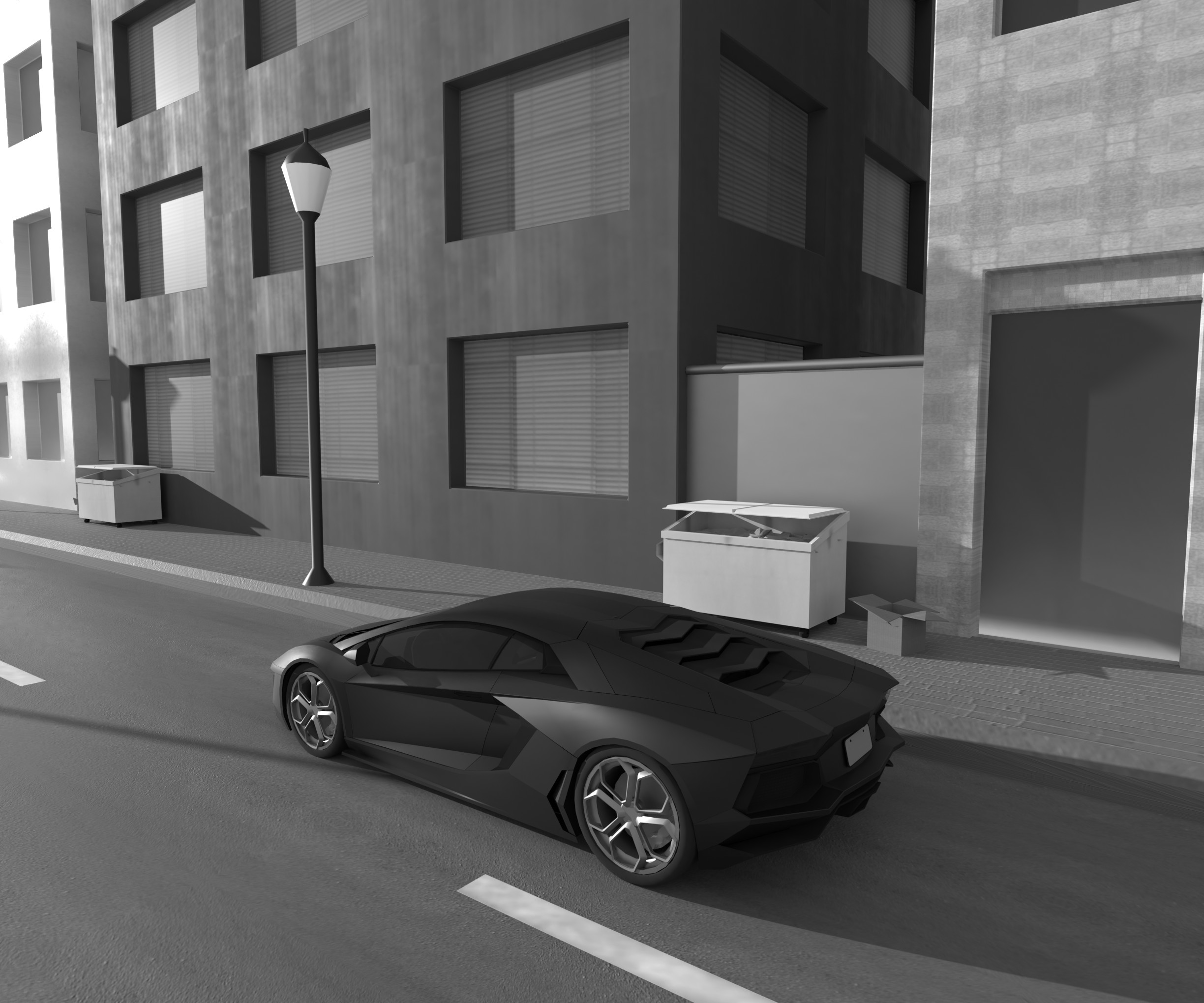}};
\end{tikzpicture}

%% file: figures/evaluation/evaluation.tikz
\begin{tikzpicture}[x=1cm, y=-1cm]
			\begin{scope}[shift={(0, 0)}]
				\node[inner sep=0pt, anchor=west] (rgb_2) at (0,-0.15) {\includegraphics[width=.23\textwidth, trim=0 100 0 0, clip]{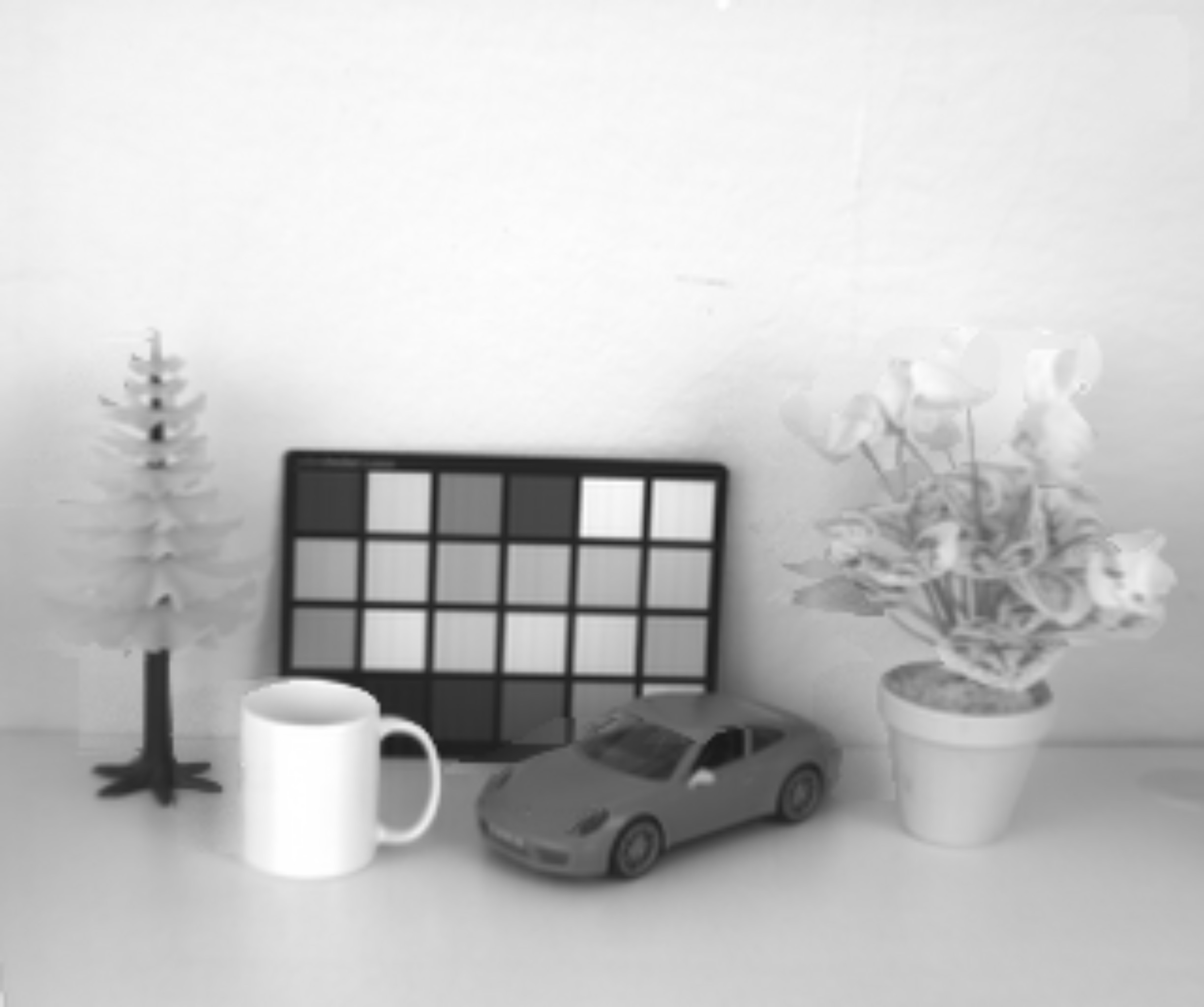}};

				\draw[draw=cyan, line width=0.25mm] (0.3, -0.6) rectangle ++(0.45, 0.68);
				\draw[draw=cyan, line width=0.25mm] (1.2, -1.715) rectangle ++(0.9, 1.425);
				\node[inner sep=0pt, anchor=west] (rgb_3) at (1.2,-1) {\includegraphics[width=.05\textwidth, trim=190 950 2010 700, clip]{evaluation/BIC}};

				\draw[draw=white!60!red, line width=0.25mm] (3, 0.23) rectangle ++(0.6, 0.6);
				\draw[draw=white!60!red, line width=0.25mm] (2.8, -1.715) rectangle ++(1.08, 1.07);
				\node[inner sep=0pt, anchor=west] (rgb_3) at (2.8,-1.18) {\includegraphics[width=.06\textwidth, trim=1800 500 300 1200, clip]{evaluation/BIC}};

				\node[fill=white, fill opacity=0.8, inner sep=1pt, anchor=west] at (0.05, -1.6) {\small BIC};
			\end{scope}

			\begin{scope}[shift={(4.2, 0)}]
				\node[inner sep=0pt, anchor=west] (rgb_1) at (0,-0.15) {\includegraphics[width=.23\textwidth, trim=0 100 0 0, clip]{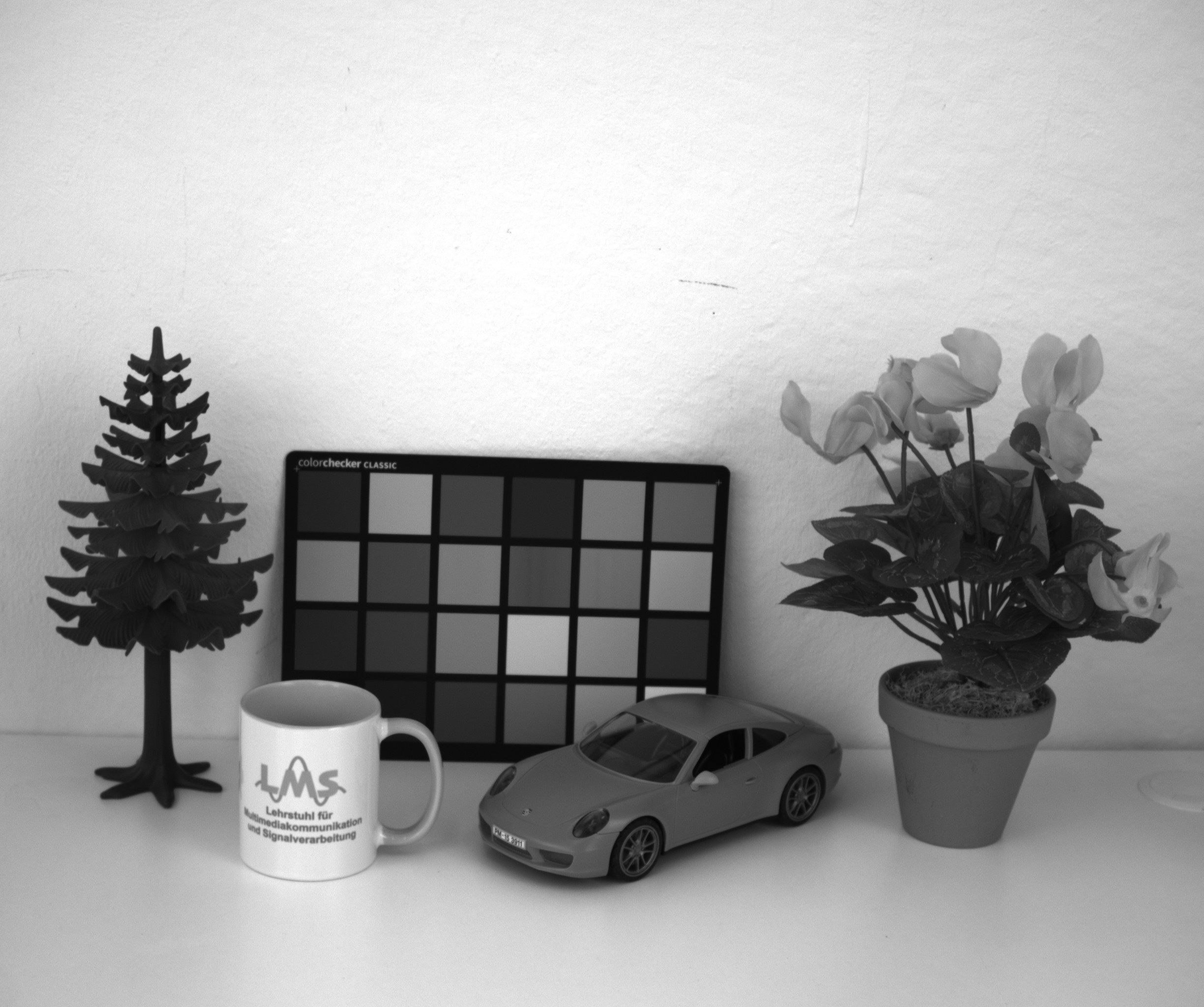}};

				\draw[draw=cyan, line width=0.25mm] (0.3, -0.6) rectangle ++(0.45, 0.68);
				\draw[draw=cyan, line width=0.25mm] (1.2, -1.715) rectangle ++(0.9, 1.425);
				\node[inner sep=0pt, anchor=west] (rgb_3) at (1.2,-1) {\includegraphics[width=.05\textwidth, trim=190 950 2010 700, clip]{evaluation/vis_high_res}};

				\draw[draw=white!60!red, line width=0.25mm] (3, 0.23) rectangle ++(0.6, 0.6);
				\draw[draw=white!60!red, line width=0.25mm] (2.8, -1.715) rectangle ++(1.08, 1.07);
				\node[inner sep=0pt, anchor=west] (rgb_3) at (2.8,-1.18) {\includegraphics[width=.06\textwidth, trim=1800 500 300 1200, clip]{evaluation/vis_high_res}};

				\node[fill=white, fill opacity=0.8, inner sep=1pt, anchor=west] at (0.05, -1.6) {\small Visible};
			\end{scope}

			\begin{scope}[shift={(0, 3.35)}]
				\node[inner sep=0pt, anchor=west] (rgb_3) at (0,-0.15) {\includegraphics[width=.23\textwidth, trim=0 100 0 0, clip]{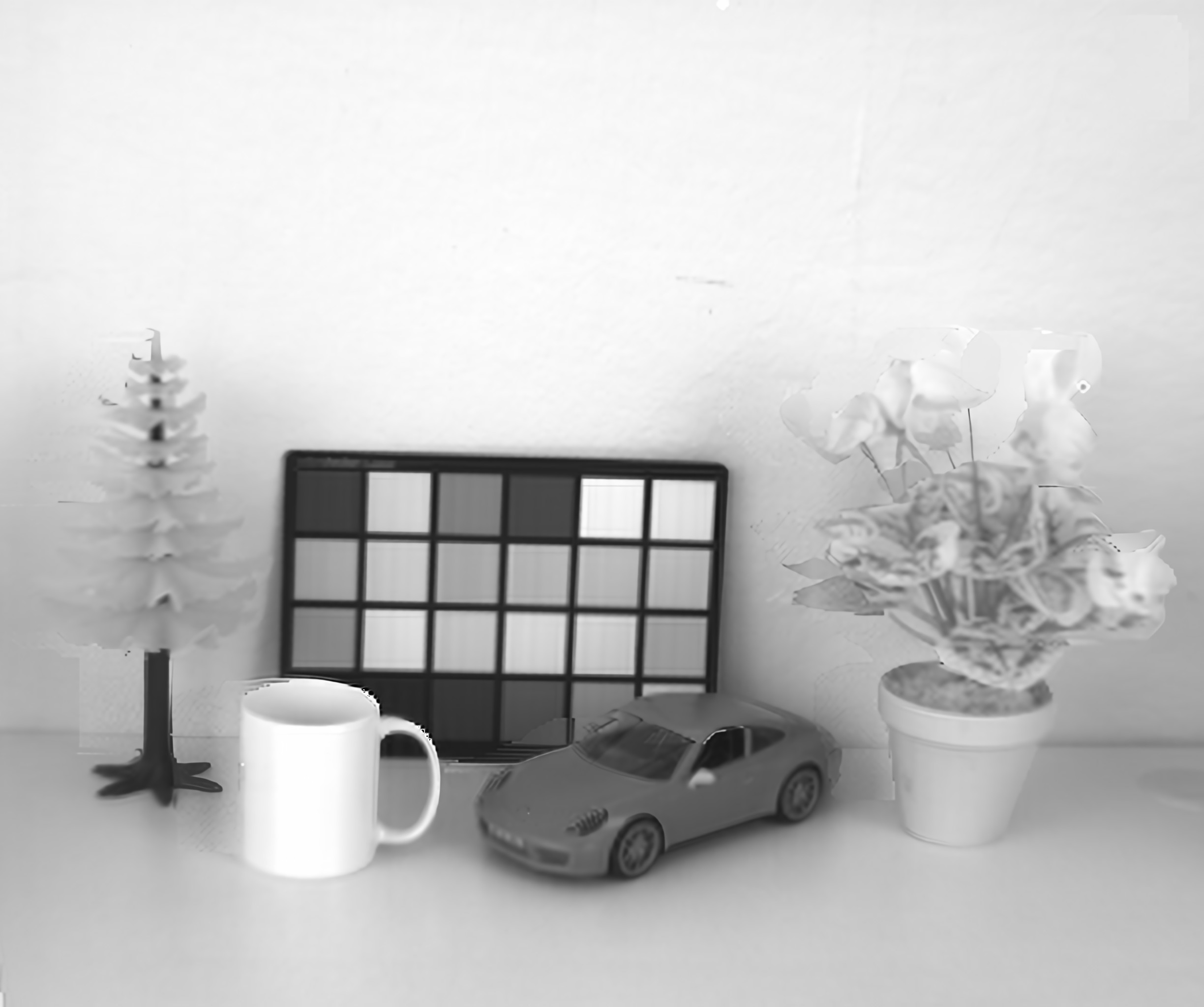}};

				\draw[draw=cyan, line width=0.25mm] (0.3, -0.6) rectangle ++(0.45, 0.68);
				\draw[draw=cyan, line width=0.25mm] (1.2, -1.715) rectangle ++(0.9, 1.425);
				\node[inner sep=0pt, anchor=west] (rgb_3) at (1.2,-1) {\includegraphics[width=.05\textwidth, trim=190 950 2010 700, clip]{evaluation/SwinIR}};

				\draw[draw=white!60!red, line width=0.25mm] (3, 0.23) rectangle ++(0.6, 0.6);
				\draw[draw=white!60!red, line width=0.25mm] (2.8, -1.715) rectangle ++(1.08, 1.07);
				\node[inner sep=0pt, anchor=west] (rgb_3) at (2.8,-1.18) {\includegraphics[width=.06\textwidth, trim=1800 500 300 1200, clip]{evaluation/SwinIR}};

				\node[fill=white, fill opacity=0.8, inner sep=1pt, anchor=west] at (0.05, -1.6) {\small SwinIR};
			\end{scope}

			\begin{scope}[shift={(4.2, 3.35)}]
				\node[inner sep=0pt, anchor=west] (rgb_3) at (0,-0.15) {\includegraphics[width=.23\textwidth, trim=0 100 0 0, clip]{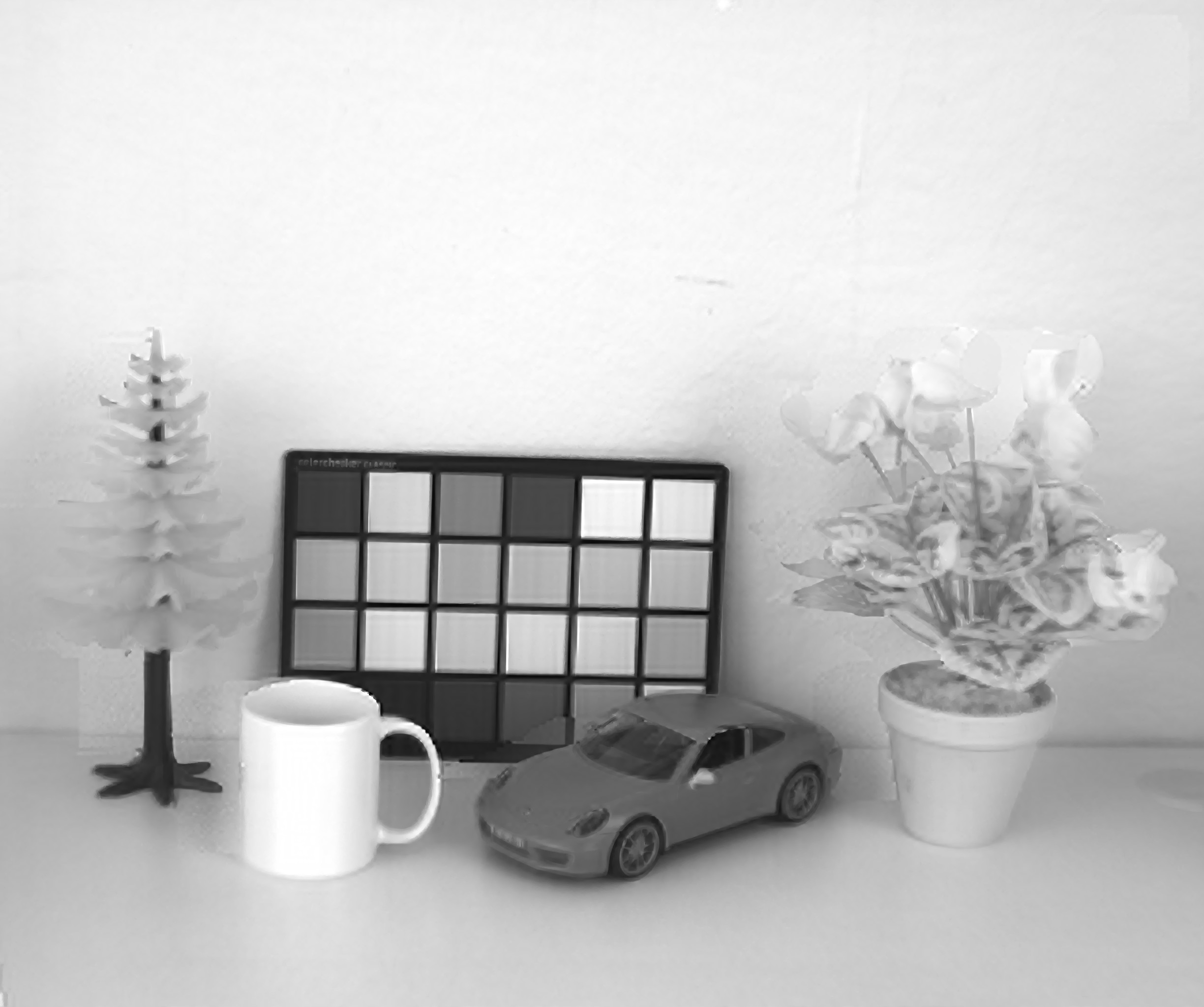}};

				\draw[draw=cyan, line width=0.25mm] (0.3, -0.6) rectangle ++(0.45, 0.68);
				\draw[draw=cyan, line width=0.25mm] (1.2, -1.715) rectangle ++(0.9, 1.425);
				\node[inner sep=0pt, anchor=west] (rgb_3) at (1.2,-1) {\includegraphics[width=.05\textwidth, trim=190 950 2010 700, clip]{evaluation/GAD}};

				\draw[draw=white!60!red, line width=0.25mm] (3, 0.23) rectangle ++(0.6, 0.6);
				\draw[draw=white!60!red, line width=0.25mm] (2.8, -1.715) rectangle ++(1.08, 1.07);
				\node[inner sep=0pt, anchor=west] (rgb_3) at (2.8,-1.18) {\includegraphics[width=.06\textwidth, trim=1800 500 300 1200, clip]{evaluation/GAD}};

				\node[fill=white, fill opacity=0.8, inner sep=1pt, anchor=west] at (0.05, -1.6) {\small GAD};
			\end{scope}

			\begin{scope}[shift={(0, 6.7)}]
				\node[inner sep=0pt, anchor=west] (rgb_3) at (0,-0.15) {\includegraphics[width=.23\textwidth, trim=0 100 0 0, clip]{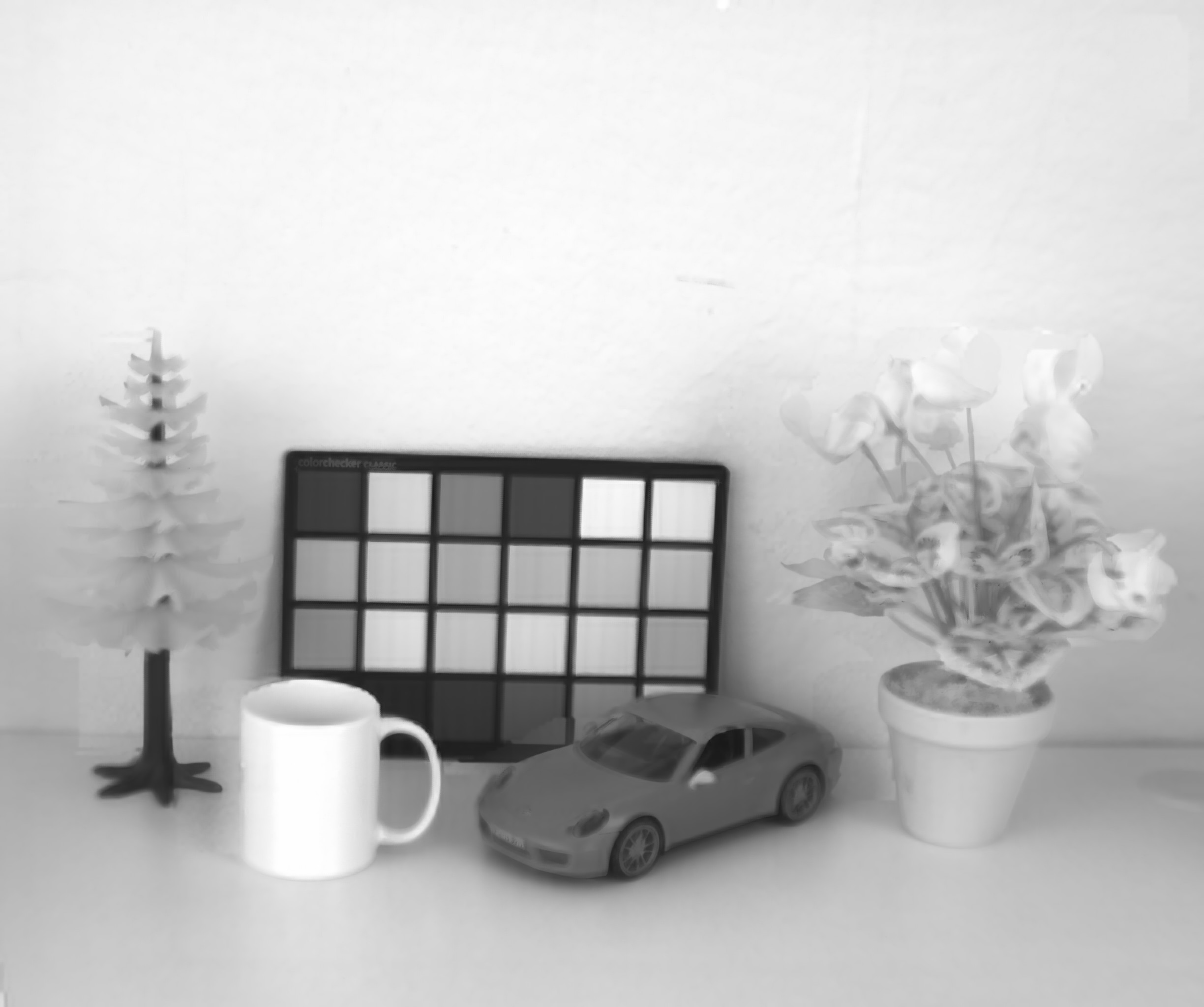}};

				\draw[draw=cyan, line width=0.25mm] (0.3, -0.6) rectangle ++(0.45, 0.68);
				\draw[draw=cyan, line width=0.25mm] (1.2, -1.715) rectangle ++(0.9, 1.425);
				\node[inner sep=0pt, anchor=west] (rgb_3) at (1.2,-1) {\includegraphics[width=.05\textwidth, trim=190 950 2010 700, clip]{evaluation/FDKN}};

				\draw[draw=white!60!red, line width=0.25mm] (3, 0.23) rectangle ++(0.6, 0.6);
				\draw[draw=white!60!red, line width=0.25mm] (2.8, -1.715) rectangle ++(1.08, 1.07);
				\node[inner sep=0pt, anchor=west] (rgb_3) at (2.8,-1.18) {\includegraphics[width=.06\textwidth, trim=1800 500 300 1200, clip]{evaluation/FDKN}};

				\node[fill=white, fill opacity=0.8, inner sep=1pt, anchor=west] at (0.05, -1.6) {\small FDKN};
			\end{scope}

			\begin{scope}[shift={(4.2, 6.7)}]
				\node[inner sep=0pt, anchor=west] (rgb_3) at (0,-0.15) {\includegraphics[width=.23\textwidth, trim=0 100 0 0, clip]{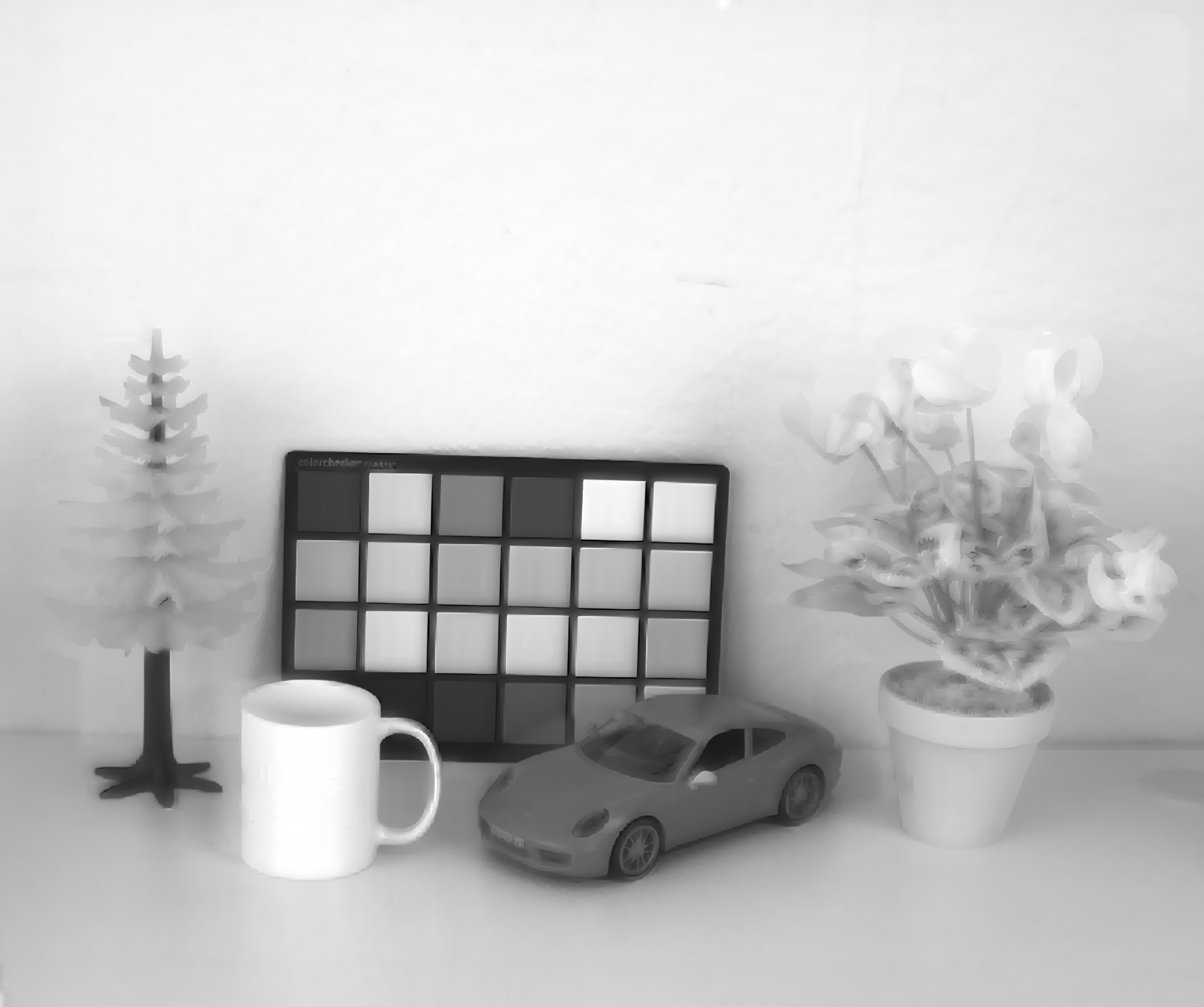}};

				\draw[draw=cyan, line width=0.25mm] (0.3, -0.6) rectangle ++(0.45, 0.68);
				\draw[draw=cyan, line width=0.25mm] (1.2, -1.715) rectangle ++(0.9, 1.425);
				\node[inner sep=0pt, anchor=west] (rgb_3) at (1.2,-1) {\includegraphics[width=.05\textwidth, trim=190 950 2010 700, clip]{evaluation/GUNet}};

				\draw[draw=white!60!red, line width=0.25mm] (3, 0.23) rectangle ++(0.6, 0.6);
				\draw[draw=white!60!red, line width=0.25mm] (2.8, -1.715) rectangle ++(1.08, 1.07);
				\node[inner sep=0pt, anchor=west] (rgb_3) at (2.8,-1.18) {\includegraphics[width=.06\textwidth, trim=1800 500 300 1200, clip]{evaluation/GUNet}};

				\node[fill=white, fill opacity=0.8, inner sep=1pt, anchor=west] at (0.05, -1.6) {\small GUNet};
			\end{scope}
\end{tikzpicture}

%% file: paper.bbl
\begin{thebibliography}{10}

\bibitem{plant}
Matheus Cardim Ferreira~Lima, Anne Krus, Constantino Valero, Antonio
  Barrientos, Jaime del Cerro, and Juan~Jesús Roldán-Gómez,
\newblock ``Monitoring plant status and fertilization strategy through
  multispectral images,''
\newblock {\em Sensors}, vol. 20, no. 2, 2020.

\bibitem{burn}
Michael~G. Sowa, Lorenzo Leonardi, Jeri~R. Payette, K.~M. Cross, Manuel Gomez,
  and Joel Fish,
\newblock ``Classification of burn injuries using near-infrared spectroscopy,''
\newblock {\em Journal of Biomedical Optics}, vol. 11, no. 5, pp. 054002, 2006.

\bibitem{plastics}
Monica Moroni, Alessandro Mei, Alessandra Leonardi, Emanuela Lupo, and
  Floriana~La Marca,
\newblock ``{PET} and {PVC} {Separation} with {Hyperspectral} {Imagery},''
\newblock {\em Sensors}, vol. 15, no. 1, pp. 2205--2227, Jan. 2015.

\bibitem{haze}
Marc~P. Hansen and Douglas~S. Malchow,
\newblock ``{Overview of SWIR detectors, cameras, and applications},''
\newblock in {\em Thermosense XXX}. International Society for Optics and
  Photonics, 2008, vol. 6939, p. 69390I, SPIE.

\bibitem{swir_noise}
Michael MacDougal, Jon Geske, Chad Wang, Shirong Liao, Jonathan Getty, and Alan
  Holmes,
\newblock ``{Low dark current InGaAs detector arrays for night vision and
  astronomy},''
\newblock in {\em Infrared Technology and Applications XXXV}, Bj{\o}rn~F.
  Andresen, Gabor~F. Fulop, and Paul~R. Norton, Eds. International Society for
  Optics and Photonics, 2009, vol. 7298, p. 72983F, SPIE.

\bibitem{camsi}
Nils Genser, Jürgen Seiler, and André Kaup,
\newblock ``Camera array for multi-spectral imaging,''
\newblock {\em IEEE Transactions on Image Processing}, vol. 29, pp. 9234--9249,
  2020.

\bibitem{gad}
Nando Metzger, Rodrigo~Caye Daudt, and Konrad Schindler,
\newblock ``Guided depth super-resolution by deep anisotropic diffusion,''
\newblock in {\em Proceedings of the IEEE/CVF Conference on Computer Vision and
  Pattern Recognition (CVPR)}, June 2023, pp. 18237--18246.

\bibitem{fdkn}
Beomjun Kim, Jean Ponce, and Bumsub Ham,
\newblock ``Deformable kernel networks for joint image filtering,''
\newblock in {\em International Journal of Computer Vision volume}, 2021, vol.
  129, pp. 579--600.

\bibitem{graph_sr}
Riccardo de~Lutio, Alexander Becker, Stefano D'Aronco, Stefania Russo, Jan~D.
  Wegner, and Konrad Schindler,
\newblock ``Learning graph regularisation for guided super-resolution,''
\newblock in {\em Proceedings of the IEEE/CVF Conference on Computer Vision and
  Pattern Recognition (CVPR)}, June 2022, pp. 1979--1988.

\bibitem{unet}
Olaf Ronneberger, Philipp Fischer, and Thomas Brox,
\newblock ``U-net: Convolutional networks for biomedical image segmentation,''
\newblock in {\em Medical Image Computing and Computer-Assisted Intervention --
  MICCAI 2015}, Nassir Navab, Joachim Hornegger, William~M. Wells, and
  Alejandro~F. Frangi, Eds., Cham, 2015, pp. 234--241, Springer International
  Publishing.

\bibitem{resnet}
Kaiming He, Xiangyu Zhang, Shaoqing Ren, and Jian Sun,
\newblock ``Deep residual learning for image recognition,''
\newblock in {\em 2016 IEEE Conference on Computer Vision and Pattern
  Recognition (CVPR)}, 2016, pp. 770--778.

\bibitem{implicit_function_theorem}
Krisorn Jittorntrum,
\newblock ``An implicit function theorem,''
\newblock {\em Journal of Optimization Theory and Applications volume}, vol.
  25, no. 4, pp. 575--577, 1978.

\bibitem{places}
Bolei Zhou, Agata Lapedriza, Aditya Khosla, Aude Oliva, and Antonio Torralba,
\newblock ``Places: A 10 million image database for scene recognition,''
\newblock {\em IEEE Transactions on Pattern Analysis and Machine Intelligence},
  vol. 40, no. 6, pp. 1452--1464, 2018.

\bibitem{dgnet}
Frank Sippel, J\"{u}rgen Seiler, and Andr\'{e} Kaup,
\newblock ``Cross spectral image reconstruction using a deep guided neural
  network,''
\newblock in {\em 2023 IEEE International Conference on Image Processing
  (ICIP)}, 2023, pp. 226--230.

\bibitem{cave}
Fumihito Yasuma, Tomoo Mitsunaga, Daisuke Iso, and Shree~K. Nayar,
\newblock ``Generalized assorted pixel camera: Postcapture control of
  resolution, dynamic range, and spectrum,''
\newblock {\em IEEE Transactions on Image Processing}, vol. 19, no. 9, pp.
  2241--2253, 2010.

\bibitem{hyvid}
Frank Sippel, J\"{u}rgen Seiler, and Andr\'{e} Kaup,
\newblock ``Synthetic hyperspectral array video database with applications to
  cross-spectral reconstruction and hyperspectral video coding,''
\newblock {\em J. Opt. Soc. Am. A}, vol. 40, no. 3, pp. 479--491, Mar. 2023.

\bibitem{vdsr}
Jiwon Kim, Jung~Kwon Lee, and Kyoung~Mu Lee,
\newblock ``Accurate image super-resolution using very deep convolutional
  networks,''
\newblock in {\em Proceedings of the IEEE Conference on Computer Vision and
  Pattern Recognition (CVPR)}, June 2016.

\bibitem{han}
Ben Niu, Weilei Wen, Wenqi Ren, Xiangde Zhang, Lianping Yang, Shuzhen Wang,
  Kaihao Zhang, Xiaochun Cao, and Haifeng Shen,
\newblock ``Single image super-resolution via a holistic attention network,''
\newblock in {\em Computer Vision -- ECCV 2020}, Cham, 2020, pp. 191--207,
  Springer International Publishing.

\bibitem{swinir}
Jingyun Liang, Jiezhang Cao, Guolei Sun, Kai Zhang, Luc Van~Gool, and Radu
  Timofte,
\newblock ``{SwinIR}: Image restoration using swin transformer,''
\newblock in {\em Proceedings of the IEEE/CVF International Conference on
  Computer Vision (ICCV) Workshops}, Oct. 2021, pp. 1833--1844.

\end{thebibliography}
